\documentclass[sigconf]{acmart}
\usepackage{subcaption,listings}
\AtBeginDocument{%
  }

\setcopyright{acmlicensed}
\copyrightyear{2018}
\acmYear{2018}
\acmDOI{XXXXXXX.XXXXXXX}

\acmConference[Conference acronym 'XX]{Make sure to enter the correct
  conference title from your rights confirmation emai}{June 03--05,
  2018}{Woodstock, NY}
\acmISBN{978-1-4503-XXXX-X/18/06}

\begin{document}

\title{ Augmented Body Communicator: Enhancing daily body expression for people with upper limb limitations through LLM and a robotic arm}

\author{Songchen Zhou}
\orcid{0009-0004-5280-2387}
\affiliation{%
  \institution{Keio University Graduate School of Media Design}
  \country{Japan}
}
\email{zhousongchen@kmd.keio.ac.jp}

\author{Mark Armstrong}

\affiliation{%
  \institution{Keio University Graduate School of Media Design}
  \country{Japan}
}
\email{mark@keio.jp}

\author{Giulia Barbareschi}

\affiliation{%
  \institution{Keio University Graduate School of Media Design}
  \country{Japan}
}
\email{barbareschi@kmd.keio.ac.jp}

\author{Toshihiro Ajioka}

\affiliation{%
  \institution{Keio University Graduate School of Media Design}
  \country{Japan}
}
\email{ajioka@kmd.keio.ac.jp}

\author{Zheng Hu}
\affiliation{%
  \institution{Keio University Graduate School of Media Design}
  \country{Japan}
}
\email{huzheng@kmd.keio.ac.jp}

\author{Ryoichi Ando}

\affiliation{%
  \institution{Keio University Graduate School of Media Design}
  \country{Japan}
}
\email{andoryoichi@kmd.keio.ac.jp}

\author{Kentaro	Yoshifuji}
\affiliation{%
  \institution{Ory Lab Inc.}
  \country{Japan}
}
\email{	ory@orylab.com}

\author{Masatane Muto}
\affiliation{%
  \institution{WITH ALS}
  \country{Japan}
}
\email{	masatane.muto@withals.net}

\author{Kouta Minamizawa}
\affiliation{%
  \institution{Keio University Graduate School of Media Design}
  \country{Japan}
}
\email{kouta@kmd.keio.ac.jp}
\renewcommand{\shortauthors}{Trovato et al.}

\begin{abstract}
  Individuals with upper limb movement limitations face difficulties interacting with others. Currently, the use of robotic arms is focused on addressing functional tasks, but there is still significant room for exploration around enhancing users’ body language capabilities during social interactions. This paper proposes an Augmented Body Communicator system that combines robotic arms and a large language model. Thanks to the incorporation of kinetic memory disabled users and their supporters can collaboratively create actions for the robot arm. The LLM system then provides recommendations about which action is most appropriate based on contextual cues in interactions with others. The system underwent in-depth user testing with 6 participants with conditions affecting upper limb mobility. Results indicate that the system enhances users’ ability to express themselves. Based on our findings we propose recommendations for robotic arms able to support disabled individuals with body language capabilities and functional tasks.
\end{abstract}


\begin{teaserfigure}
  \includegraphics[width=\textwidth]{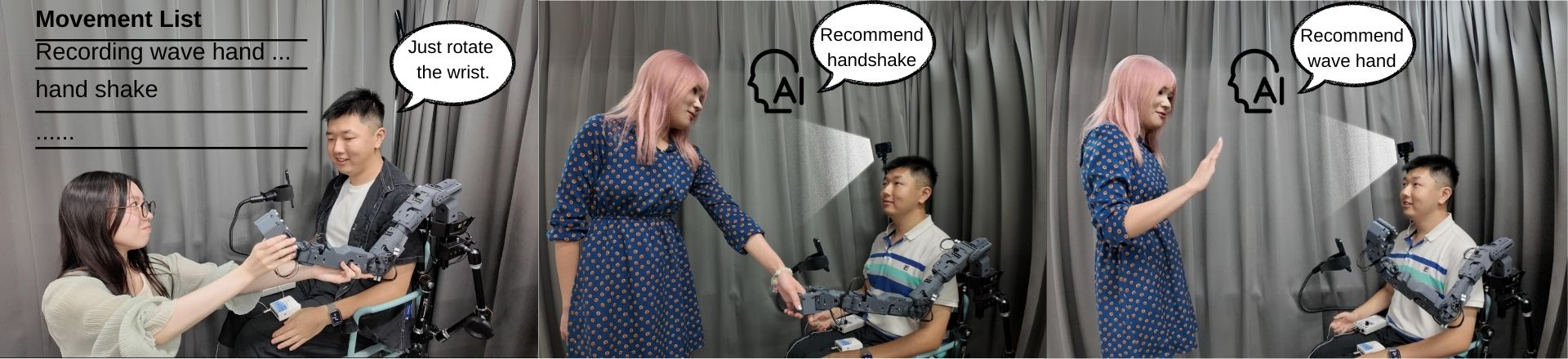}
  \caption{With the assistant's help recording movements, AI recommends robotic arm actions like handshake and wave in response to interactions captured by a camera. }
  \Description{A user asks an assistant to help record robotic arm movements. While others are interacting with the user by shaking hands or greeting, a camera captures photos of the interaction partner, and an LLM AI recommends the user to perform handshake or wave hand actions with the robotic arm in response.}
  \label{fig:teaser}
\end{teaserfigure}

\received{20 February 2007}
\received[revised]{12 March 2009}
\received[accepted]{5 June 2009}

\maketitle

\section{Introduction}
The causes of upper limb movement difficulties are diverse, including amputation, neurological disorders, muscular diseases, skeletal issues, congenital deformities, and arthritis \cite{walker2004prevalence, d2013clinical, dubuisson2017disease, furlan2011assessment, raghavan2015upper}. As a result of loss of mobility, individuals' daily lives are significantly impacted, particularly in relation to the performance of self-care functional tasks \cite{raghavan2015upper, boyd2001management, hudak1996development, roh2013clinical}, but also in their ability to express emotions through body language \cite{de2009bodies, baylor2011qualitative, hartley2015spinal, purton2021experience}.

For disabled and non-disabled people alike, nonverbal communication often conveys emotions and attitudes that words alone cannot fully express, playing a crucial role in ensuring accurate understanding in social interactions which can affect personal wellbeing \cite{Nonverbalcommunication, meeren2005rapid, laureys2005locked}. Upper limb movements are a significant component of body language expression including conveying specific emotions and performing socially meaningful gestures \cite{witkower2019bodily, wang2015adaptive, piana2016adaptive, bernardis2006speech}. Research indicates that robots can effectively convey emotions through body language in a manner that most people can accurately identify \cite{beck2012emotional}.

Existing research on how technology can support people with severe upper limb mobility often focuses on helping individuals regain functionality \cite{chu2018robot,kieliba2021robotic,proietti2023restoring}. Additionally, there have been studies exploring the use of extra robotic arms to support self-feeding \cite{handelman2022shared, parikh2024design}, picking up and placing objects \cite{lee2009brain, jain2015assistive, house2009voicebot}, or interacting with one's physical surrounding \cite{maheu2011evaluation, petrich2021assistive}. However, research on the use of robotic arms to augment body language for those with severe mobility restrictions is scarce \cite{goldstein2002correlates, orejuela2019self, valencia2021co}. 

In a co-design workshop for socially assistive "sidekicks" able to augment non-verbal communication for Augmentative and Alternative Communication devices (AAC) users, Valencia et al.\cite{valencia2021co} documented how such artifacts could help disabled people express meaningful cues through movement to supplement or substitute verbal exchanges with conversation partners. A recent study on a series of live performances involving a disabled artist has also shown how the use of robotic arms controlled via Brain-Computer Interface can help strengthen the sense of connection between the artist and the audience\cite{B2J}. However, such systems are still in the early stage of conceptualization or deployed in extremely specific settings, which don't necessarily match daily life requirements.

To address these issues, we have developed an Augmented Body Communicator (ABC) system featuring a 7 DoFs robotic arm with kinetic memory combined with a camera system incorporating Large Language Model (LLM) AI image recognition to help recommend contextually appropriate movements via a browser base application compatible with various input modalities. The ABC system allows users with upper limb disabilities to record robotic arm movements with the help of a chosen assistant to fulfill the user's unique bodily expression needs. The ABC control system is built on a web-based application, enabling users to control it using a phone or computer with auxiliary devices. The control system allows for two different selection modes. The first is a basic analogue selection that enables the user to search for a desired movement based on the assigned name or most recent use. The second integrates an external camera that the user can utilize to take a picture when the person they are interacting with makes a particular gesture, the ABC system uses an LLM AI image recognition model to interpret the gesture and then recommends a suitable response movement for the user based on their personalized repository of recorded movements. After receiving the recommended options, the user can quickly activate the robotic arm for interaction.

We evaluated the accuracy of recognition of the ABC system based on images of different quality as well as the accuracy of recommendation compared to participants' expectations. Additionally, 4 users with muscular dystrophy and 2 users with Amyotrophic Lateral Sclerosis (ALS) were invited for in-depth user testing. The evaluation, conducted through questionnaires, interviews, and observations, indicated that this system significantly enhanced users' daily expressive abilities and enabled them to perform movements and gestures that were either impossible or extremely difficult for them. Participants valued the ability to use the ABC to participate and initiate common social exchanges, communicating in a more personal manner with their loved one, or simply scratch an itch they were unable to reach. Functional challenges still remain including the fact that, albeit short, the processing time still creates some delays in social interactions and how the AI-powered recommendation requires conversational partners to make a gesture to be able to suggest potential responses. Based on our results, we formulate recommendations around the future development of ABC systems that leverage robotic arms for supporting social and functional interactions in daily life for people with restricted upper limb mobility.
Our contributions can be summarized as follows:
\begin{enumerate}
    \item The development of the first ABC system integrating a robotic arm with kinetic memory featuring a control system supporting analogue choices and LLM AI-powered image recognition recommendations
    \item Insights gathered from people with disabilities and their conversational partners about the effectiveness and limitations of ABC in social settings
    \item Design implication for the development of ABC systems supporting people with upper limb mobility limitations and the potential role of AI for these applications
\end{enumerate}

\section{Related Work}
\subsection{Augmentative and Alternative Communication} \label{AAC}
For people who face significant challenges in their ability to communicate AACs represent an effective tool to support self-expression and social interaction \cite{valencia2020conversational, kane2017times}. The most commonly used AAC tool is the transparent text board, which uses letters, pictures, or symbols to facilitate communication. Caregivers rotate the board and identify the text by pointing, reading aloud, and observing the user’s facial expressions, while users transmit their commands by pointing or gazing at the content \cite{zangari1994augmentative, vanderheiden2003journey}. With the development and increased availability of technology, different evolutions of these traditional boards have been created, integrating interface modalities ranging from switches \cite{wiegand2014rsvp, elsahar2019augmentative}, gaze-based input \cite{ball2010eye, fiannaca2017aacrobat}, brain-machine interfaces \cite{brumberg2018brain, peters2022systematic}, and more recently generative AI models \cite{valencia2023less, cai2023using}, to accommodate different input capabilities and allow for independent communication without the need for an external assistant. 

In recent years, researchers and developers have also explored different ways to convey important information beyond the content of a verbal message. Strategies have included, among others, the integration of expressive voices \cite{fiannaca2018voicesetting}, awareness displays facing the AAC user's conversational partner \cite{sobel2017exploring}, and mobile-based smart badges \cite{curtis2024breaking}. Beyond visual displays and auditory clues, we found two notable examples of recent research that focused on co-developing expressive objects to support AAC users with non-verbal communication through motion \cite{valencia2021co, valencia2021aided}. However, these proposed "sidekick" interfaces which use robot motions and gestures for nonverbal communication have relatively limited motion possibilities, are largely intended to be used as support for verbal AAC communication, and can require the creation of individualized add-ons potentially placing a customization burden on the user.

We propose that the creation of a more complex and flexibly programmable robotic interface can support more substantial non-verbal communication for users with restricted mobility, becoming an independent means of expression, rather than solely a supporting agent for verbal exchanges. We believe that this is more aligned with how individuals use their bodies, and particularly their arms, as communicative interfaces in everyday social exchanges.

\subsection{Robot for Body Expression}

Scholars in the field of Human-Robot Interaction have carried out substantial research to investigate how robotic agents with different form factors can communicate their intentions non-verbally through movements \cite{nakata1998expression, urakami2023nonverbal, van2022nonverbal}. This included studies specifically focusing on the expressiveness of robotic arms \cite{chatterjee2022body, kumar2024exploratory}. 

Meanwhile, the use of robotic arms as a means to augment bodily expression has, so far been predominantly focused on artistic applications, particularly involving non-disabled individuals. Abrahamsson delves into Stelarc's philosophical and artistic challenges to traditional concepts of the self and the body's limits, portraying the body as a zombie-cyborg hybrid intrinsically linked with technology \cite{abrahamsson2007conversation}. Ladenheim et al. present "Babyface," an artistic critique of feminine gender performance through a wearable robotic device, challenging assumptions about the female ideal in technology \cite{ladenheim2020live}. Similarly, Yamamura introduces JIZAI ARMS, a wearable robotic limb system with detachable arms designed for social interaction \cite{yamamura2023social}.

In contrast, research focusing on the use of robotic arms by individuals with disabilities has primarily focused on functional task completion, such as self-feeding \cite{handelman2022shared}, assisting with bathing \cite{zlatintsi2020support}, dressing \cite{kapusta2019personalized}\cite{canal2019adapting}, or object manipulation and environmental interaction \cite{maheu2011evaluation, petrich2021assistive}. Examples of research investigating how disabled operators leverage nonverbal communication via robotic interfaces have been instead centred on telepresence avatars which represent the individual in a different location, rather than augmenting bodily expressions in situ \cite{Rode2018, barbareschi2023both}. The only exception we were able to find was the recent work documenting the design and deployment of a Brain Machine Interface-controlled system to augment the stage presence and bodily expression of a DJ with mobility restriction during a live performance \cite{B2J}. However, the goal and context of this research are very specific to its implementation in a public concert and feature an experienced artist as the sole user. The potential of using robotic arms to support nonverbal communication for people with limited mobility in daily social interactions is still underexplored.

\subsection{AI and Large Language Models for Inclusive Verbal and Nonverbal Communication} \label{LLM}

Given their strong capabilities in natural language generation, reasoning, and image understanding LLMs have become an increasingly popular topic of research in HCI and beyond \cite{achiam2023gpt, aubin2024llms, gao2024taxonomy, xu2024can, yang2024human}. In the context of accessibility, LLMs have been leveraged to support, with varying degrees of success \cite{glazko2023autoethnographic}, tasks including non-sighted navigation \cite{yu2024human}, alternative text generation \cite{jandrey2021image}, sign language interpretation \cite{gong2024llms, papastratis2020continuous}, non-standard speech recognition \cite{ayoka2024enhancing}, and text prediction for AAC users \cite{valencia2023less,yang2023demonstration,yang2023designing}.

For AAC users systems featuring AI and LLM models have been primarily focused on reducing the burden of text generation to improve the speed and ease of communication \cite{valencia2020conversational, yang2023demonstration, yang2023designing, hackbarth2024voices, ascari2020computer}. Such systems generally take text or gesture input from the user and convert it into speech. Another field of application is that of image recognition for sign language translation to support mixed-ability conversations \cite{wen2021ai, adeyanju2021machine, ananthanarayana2021deep, bragg2021fate}. The predominant input modality for these applications is motion analysed via video recording with outputs including spoken audio and written text. Both these scenarios are focused primarily on supporting verbal communication from both sides, although each side might be using a different format including speech, signs or written text. In inclusive nonverbal communication, AI and LLMs have also been proposed for capturing and translating body language to support social exchanges specifically for visually impaired or autistic individuals \cite{morrison2017imagining, begel2020lessons,dantas2022recognition,qiu2020understanding}. 

Our study sought to combine these three different areas, leveraging a LLM model to identify socially meaningful gestures performed by a conversational partner and provide contextually appropriate recommendations for users of an Augmented Body Communication tool, thus reducing the burden of interaction for nonverbal communication.

\section{Design of an Augmented Body Communicator system}

The participatory design process that led to the development of the ABC system is situated in the context of a larger project involving researchers from the HCI, Robotics and Assistive Technology fields, as well as people with ALS and other conditions causing severe mobility limitations. The goal of the project is to design technology supporting better opportunities for social interaction and participation. The ABC system was initially conceptualized based on the needs and ideas of one of the members of our design team, an adult man in his late 30s who has been living with ALS for over 10 years and as a result, experiences significant mobility challenges which impacts his ability to communicate with others in everyday life. Although he is extremely proficient in the use of both analogue and digital gaze-based AACs, these devices only support his verbal communication, and he wanted to explore ways to expand possibilities for nonverbal interactions. The rest of our team includes researchers and designers with and without disabilities and who have different professional expertise ranging from robotics and engineering to rehabilitation and accessibility. Throughout the design process, we sought to engage in open collaboration leveraging our technical skills and lived experiences to explore technological, practical, and social implications which determined the characteristics of the ABC system.  In the following sections, we first present the exploratory investigation collectively carried out by the team to identify basic design requirements and constraints, then we introduce the ABC system including the robotic arm design and the control system utilized by the user.

\subsection{Exploratory Investigation and Design Concept}
To understand the contexts in which the ABC system could benefit users with mobility limitations in nonverbal communication, we conducted five observational sessions in varied settings—professional, family, public, and shopping environments. During these sessions, one or more team members accompanied our ALS collaborator to discuss her daily social interactions and evaluate the support provided by existing AAC devices. The collaborator explained her actions, demonstrated interactions with various assistive devices, expressed frustrations with current strategies, and proposed ideas for how an ABC system could add value. Meanwhile, the team asked follow‐up questions and captured notes, photos, and videos to document the scenarios. All collected material was later reviewed to inform the key design features of the ABC system.

Fieldwork revealed that users need to perform gestures such as waving goodbye, pressing elevator buttons, indicating items to purchase, greeting, affirming, negating, and shaking hands. We categorized these interactions into two groups—interaction with objects and interaction with people—with this study focusing primarily on the latter. Observations showed that when meeting or saying farewell to ALS patients, non-disabled individuals may wave, but the patients often cannot respond, leading to frustration. Furthermore, when ALS patients try to convey their intentions, they frequently struggle to express them quickly enough for others to understand, resulting in missed conversations and feelings of sadness. 

One of the first requirements discussed was to ensure that the designed system, whatever its form factor, was compatible with the use of different platforms and input interfaces. This was motivated by the observation that the collaborator with ALS employs various devices such as a Piezo-pneumatic sensor (PPS) switch, the iPhone switch control function, and the Tobii Dynavox Eye Mobile Plus eye-tracking device with either a tablet or a PC in different situations. These choices were largely motivated by contextual preferences, such as using a PC with eye tracking and a PPS switch at work for greater power and accuracy, and an iPhone with a PPS control device in daily life because of its portability. Such variability drove our decision to implement a control system which operates via a web browser to accommodate changes between devices and support compatibility with one's preferred interface.

The second design decision revolved around the idea of leveraging the form factor of a robotic arm for the ABC system. This was motivated by reflections around the use of arms in nonverbal communication and the performance of various social interactive gestures, which would make it intuitive for one's conversational partners to understand the actions of the ABC and respond accordingly. Moreover, research on commercially available robotic arms has shown how such devices can be attached to a wheelchair relatively easily and have received positive ratings concerning both functionality and social acceptability \cite{maheu2011evaluation, beaudoin2019long}.

Observations also generated discussions around the complexity of both verbal and nonverbal social interactions including the need to account for the desire to create a large set of communicative gestures, which are specific to the individual and evolving over time. As a result, we decided to include an interface which would support users' ability to record whichever motions they wanted the robot arm to perform and store them in the system for future selection. While supporting personalized forms of nonverbal expressions, we also recognized how allowing for the recording of potentially unlimited gestures, could lead to a cumbersome selection process where the user would either need to cycle through a large number of options or have to remember and type movement names. Taking inspiration from how AI and LLMs have been used to support inclusive communication (see section \ref{LLM}), we decided to implement a system where a web-based image recognition API could be leveraged to recognize contextual clues from conversation partners to recommend appropriate robot arm movements based on previously assigned names. 

\subsection{System Implementation}
The ABC system incorporates a custom-designed robotic arm with 7 DoF and a web-based control application to support the user in recording meaningful movements and selecting them at desired times. In the following subsections, we introduce each component.

\begin{figure}[h]
    \centering
    \includegraphics[width=1\linewidth]{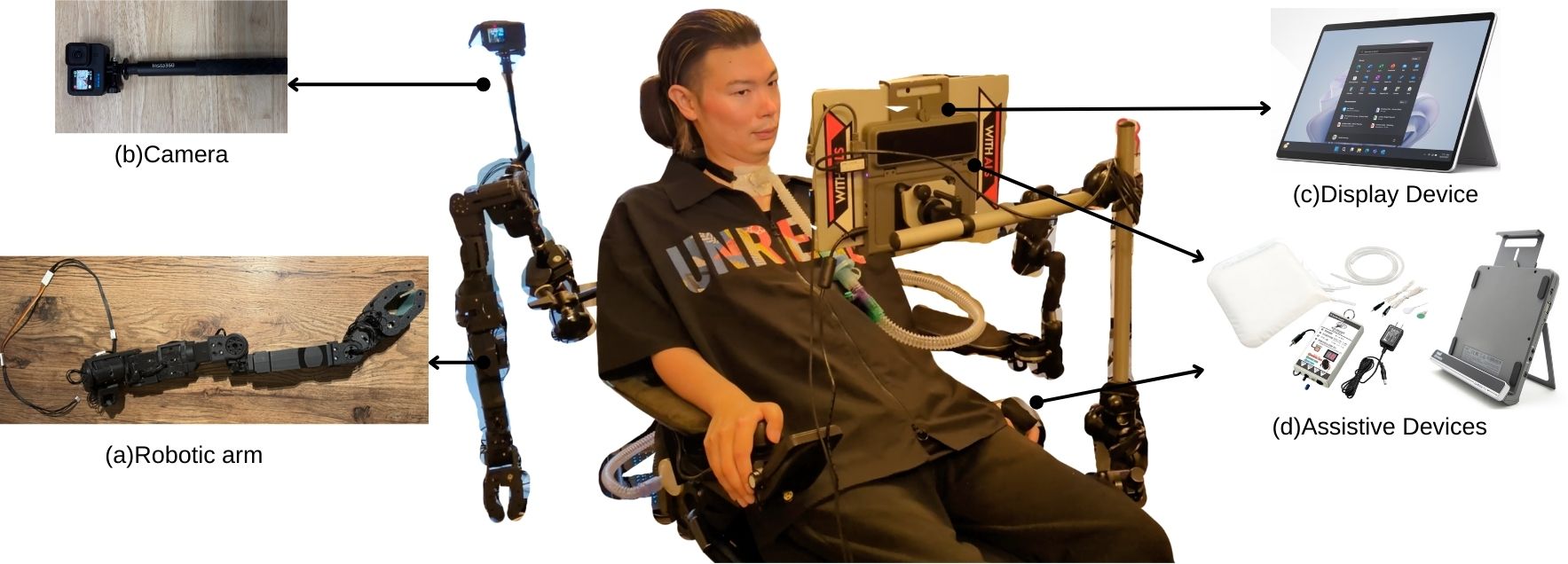}
    \caption{System Setup}
    \Description{This image shows a setup involving a robotic arm, a camera, a display, and assistive devices positioned on a chair to illustrate their arrangement.}
    \label{fig:enter-label}
\end{figure}
\subsubsection{Physical Apparatus}

The ABC system comprises five main hardware components: (1) a robotic arm mounted near the user’s shoulder to perform selected movements that augment body language, (2) a camera capturing the interaction partner’s actions in front of the user, (3) a display device presenting selectable movement options, (4) the user’s assistive device (e.g., a PPS switch or eye tracker), and (5) a laptop PC hosting the application’s web server.

We partnered with company {remove for anonymization} to develop an anthropomorphic robotic arm that enables users to physically express themselves via customizable movements. Company X employed 3D modeling and printing, allowing easy addition of features or material changes based on feedback. The arm comprises a U2D2 controller and eight servo motors (three XM540-W150 at 7.3 N·m stall torque, and five XM430-W350 at 4.1 N·m). It has 7 degrees of freedom (excluding the end effector), matching the ones of a human arm. We use DYNAMIXEL motors, which provide real-time angle data via the dedicated SDK. By controlling each joint and recording its angle, we capture the arm’s anthropomorphic movements (its “kinetic memory”). Two clamps secure the arm to the wheelchair near the user’s shoulder.

The camera is positioned at the user’s eye level on an adjustable stand, ideally with a small forward-facing screen to let the interaction partner know they are in frame (we used a GoPro Hero 10 with a minimal front screen). The user can employ any internet-capable device (e.g., phone, computer) to access the web server, view movement options for the arm, and control the system by taking a photo or selecting a movement manually. After the user clicks the photo button on the web UI using the assistive device, the system will call the LLM (gpt-4o-2024-05-13 Vision) to analyze the partner's body language and suggest appropriate actions to take.

The computer runs the robotic arm control system, calls the LLM API, and provides separate control interfaces for the user and the assistant on the same local network using different ports.

For safety, we monitored motor positions at 30 Hz and stopped the arm immediately if any joint deviated from its target position for more than 0.5 seconds. Both user and assistant interfaces also offered an emergency stop feature in their interface.  During experiments, a team member supervised the robotic arm and managed the U2D2 controller’s power switch, ensuring a quick stop whenever necessary.

\begin{figure}[h]
    \centering
    \includegraphics[width=1\linewidth]{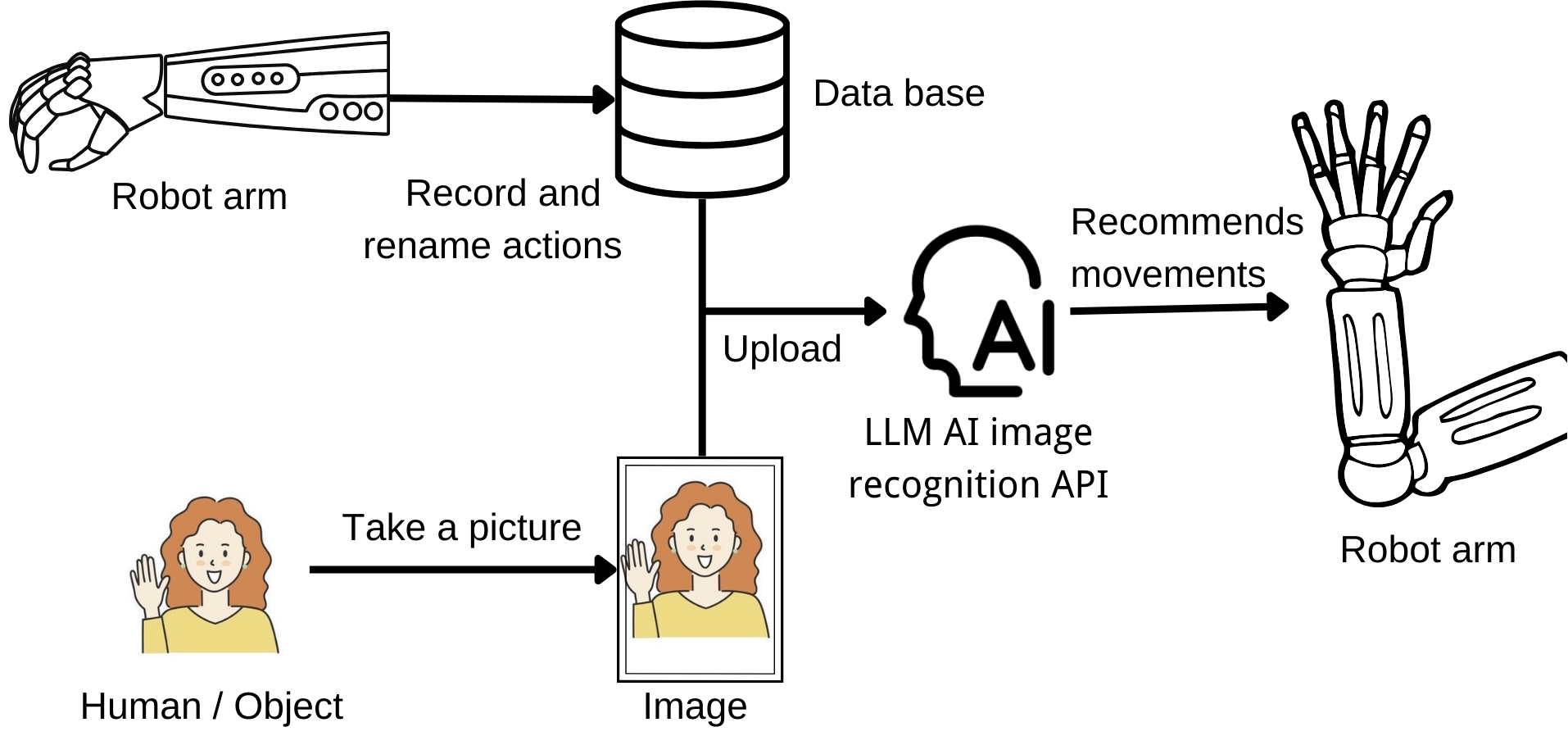}
    \caption{System Workflow}
    \Description{The diagram shows a process where a camera captures a person or object, and the LLM AI recommends response actions. The assistive robotic arm records and renames actions, storing them in a database for future recommendations. The captured image is uploaded to the LLM AI image API, which suggests corresponding robotic arm actions for the user to select and interact with others.}
    \label{fig:enter-label}
\end{figure}
\subsubsection{Control System}


We developed separate system interfaces for users with limited mobility and their assistants, enabling collaborative recording, naming, and triggering of mechanical arm movements. During daily interactions, users can search their pre-recorded movements or take a single photo, which the system analyzes to recommend suitable responses. Once a user confirms the selected movement, the mechanical arm executes it. Initially, we employed image segmentation to enhance performance, but testing showed little improvement in action recommendation accuracy. Thus, we removed this step to reduce the user’s cognitive load, speed up selection, and better support real-time interaction.


To ensure practical daily use, the device must remain power-efficient and maintain low processing times under varying network conditions. Consequently, only a single image is uploaded—rather than multiple images or a video—to prolong battery life, minimize upload volume, reduce API latency, and accelerate the interactive process. In session 4 Study 1, recommendations based on a single photo achieved high accuracy, demonstrating that the LLM can generate reliable responses from a single image. Furthermore, considering that daily interactions are replete with environmental sounds and noise, and that the spoken words accompanying body postures may not perfectly correspond to the gestures or may be produced at a distance from the user, coupled with the ALS user’s nearly lost ability to speak due to disease progression, audio was not included as part of the input.

\begin{figure}[h]
    \centering
    \includegraphics[width=1\linewidth]{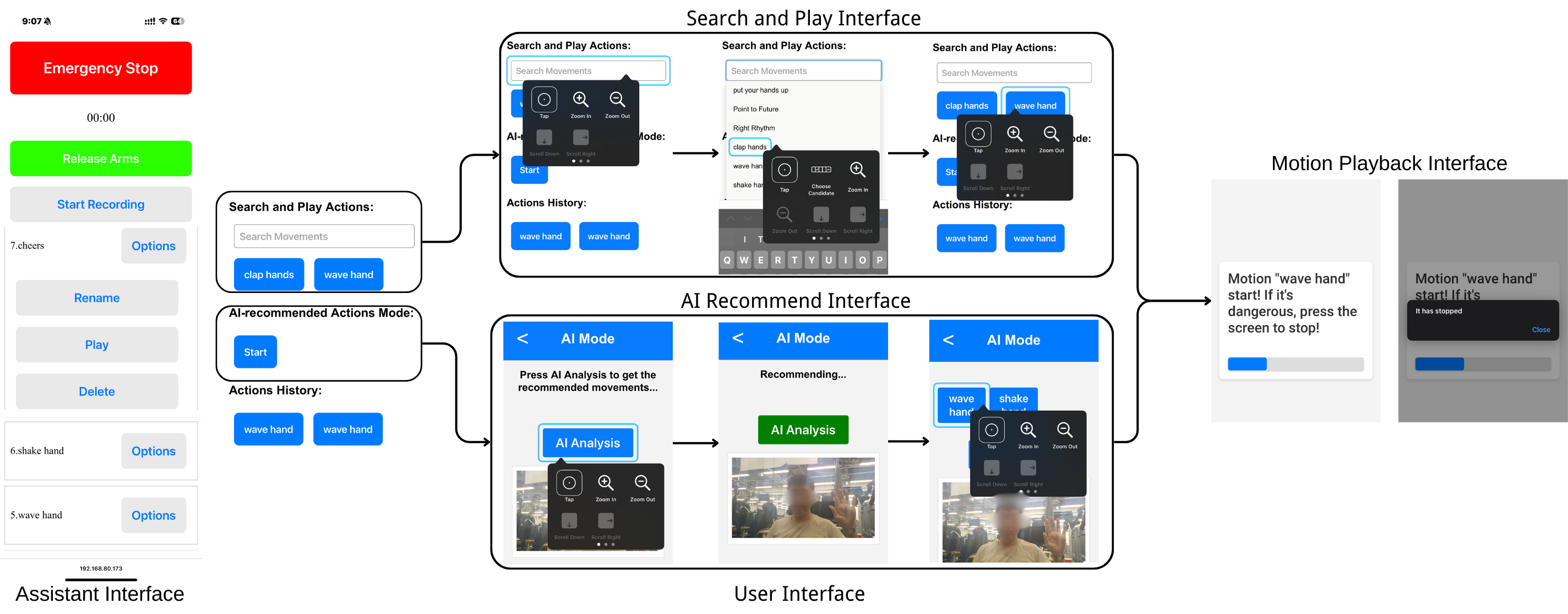}
    \caption{Assistant and User Interfaces}
    \Description{The image shows the assistant interface with buttons for "Emergency Stop," "Release Arms," "Start Recording," and options for renaming, playing, or deleting pre-recorded actions like "cheers," "shake hand," and "wave hand." 
    
    And the user interface for a system that allows users to control a robotic arm through two main modes: "Search and Play Mode" and "AI Recommend Mode."Search and Play Mode: Users can manually search for and select predefined actions (e.g., "clap hands," "wave hand"). Selected actions are displayed in the action history, and they can start motions with an emergency stop option.AI Recommend Mode: The system uses AI analysis to recommend actions based on the user's environment or posture. Once AI recommendations are generated, the user can choose and execute these actions, with an emergency stop feature also available.The emergency stop feature ensures safety by allowing users to stop any action if it becomes dangerous.}
    \label{fig:interface}
\end{figure}

\textbf{Control Interface for Assistants:} Includes an emergency stop and a button to unlock the mechanical arm, options to record and name new movements, play them back, and delete them. Assistants can lock the arm if it moves unexpectedly, unlock it after confirming safety, start and stop recordings while creating new movements, review recorded actions by playing them back, and delete those that do not meet expectations (See Figure \ref{fig:interface}.

\textbf{Control Interface for Users:} The interface has two modalities. In the manual search and play modality, users can independently search for or play back the mechanical arm movements pre-recorded with an assistant. In the AI recommended modality, the system can capture a person’s actions through a camera and apply AI-based image recognition to offer recommended responses based on the user’s stored actions. Users may also adjust the sensitivity and control settings of their assistive devices connected to the application (i.e. PPS switch or eye tracking). In case of abnormal arm movement or danger, the user can quickly stop and lock the arm by clicking anywhere on the playback interface. The lock is released once safety is assured (See Figure \ref{fig:interface}).

\section{Study 1: Accuracy testing of AI LLM system}

To evaluate the reliability of the AI LLM system we tested both accuracy of recognition (e.g. the ability to correctly identify the action of a conversational partner captured in an image), and accuracy of recommendation (e.g. the ability to select the most appropriate gesture to utilise in a specific situation based on a library of available movements). To test accuracy of recommendation we recruited 8 female and 6 male non-disabled adult participants. They were asked to score their most likely responses to the actions of a person in a photograph from a list of 12 gestures. Their responses were compared with the actions recommended by the LLM.  
To evaluate the accuracy of recognition, we tested the LLM's ability to identify specific gestures from pictures featuring different light conditions, motion artefacts, and partial occlusion.
\subsection{Participants and Procedure}
In the experiment, we collected images representing 11 common bodily interaction gestures often used in daily life. Participants were shown these images and asked to imagine how they would respond to each gesture (e.g., wave hand, thumb up, clap hands, shake hand, thumb down, raise hand, point finger, fist bump, high five, hug, OK, pat). They rated response gestures from a list of 12 options on a 5-point Likert scale. 
Since the third image in Fig. 5 can correspond to multiple response options, such as high five or raise hand, an additional option was added for both the user and LLM to choose from.
After completing the experiment, we analyzed the total scores of the gestures across 14 participants and ranked them to identify the top three responses with the highest scores to each of the gestures represented in the pictures. The LLM was then given the same set of images and response gesture names and tasked with generating gesture recommendations, based on the same list of 12 movements, ranking them based on likelihood.
\begin{lstlisting}[breaklines=true, basicstyle=\ttfamily, columns=fullflexible]
Prompt for LLM AI:
"The image depicts a scene where the user is facing the person they wish to interact with. Refer to the following action list: {recorded_actions}, as well as the scene information and the actions of the person in the image. Determine which actions the user is most likely to respond to and order them by likelihood. Return only the action names that exactly match those in the original action list {recorded_actions}, separated by commas, excluding 'init'. Suggest 2~3 suitable actions."
\end{lstlisting}


To evaluate the accuracy of LLM gesture recognition, we utilized images featuring different artifacts. These images represent eight scenarios: underexposed, normal, overexposed, high motion blur, moderate motion blur, low motion blur, partial subject, partial subject with off-focus, and interference from others. These scenarios simulate different challenges that may arise in real-world applications of gesture recognition systems. For each gesture recommendation, a new round of dialogue was initiated to prevent LLM's context memory from influencing the accuracy of the results.

\subsection{Findings}
The participants’ response ratings and the LLM’s recommended results are illustrated in the figure below:
\begin{figure}[h]
    \centering
    \includegraphics[width=1\linewidth]{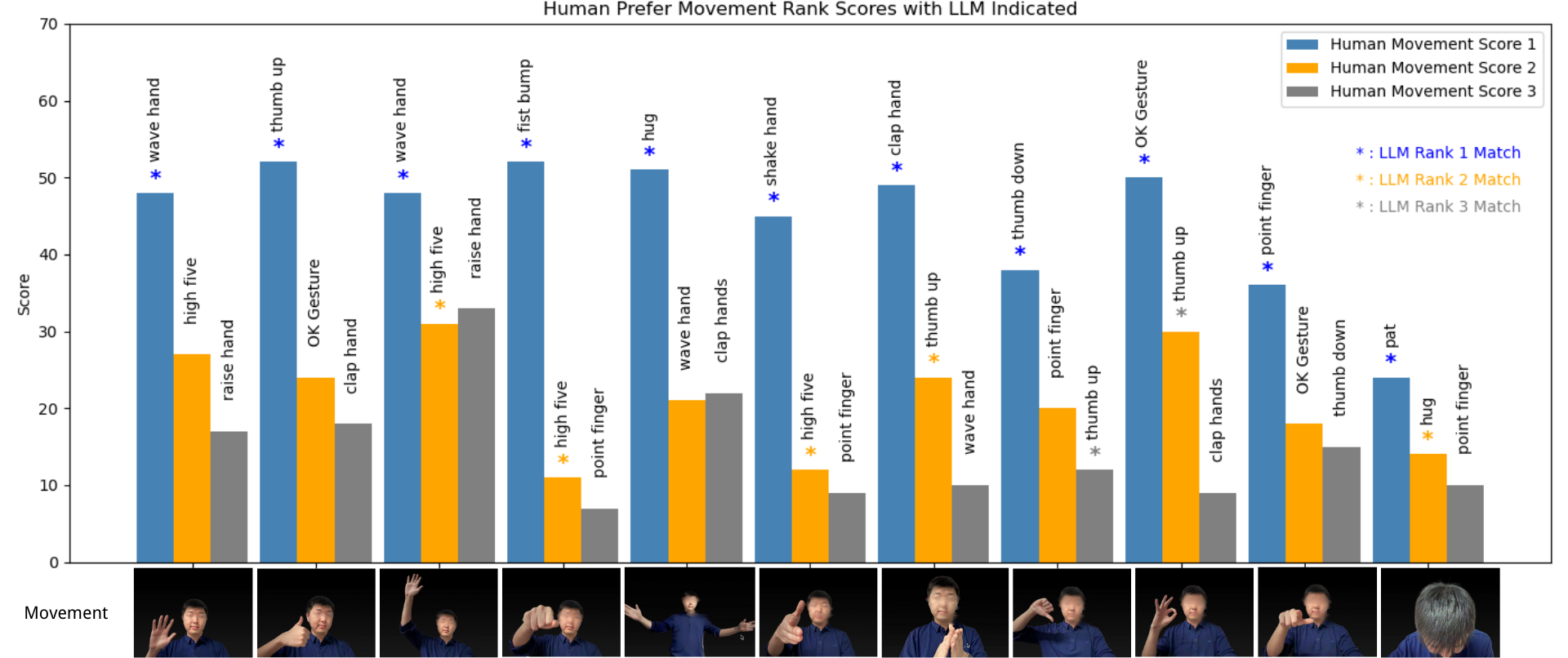}
    \caption{Human Prefer Movement Rank Scores with LLM Indicated}
    \Description{
    Figure illustrates the comparison between human preference scores and LLM recommendations for 11 common bodily interaction gestures. The X-axis represents the gestures, with images for visual reference, while the Y-axis shows scores rated by 14 participants using a 5-point Likert scale. 
    Blue, Orange, and Gray Bars** indicate the top three human-ranked gestures (Rank 1, Rank 2, Rank 3, respectively).  
    Stars denote matches between the LLM's recommendations and human rankings, with blue, orange, and gray representing Rank 1, Rank 2, and Rank 3 matches, respectively.
    
    This figure highlights the alignment between human-selected gestures and LLM predictions, showcasing a strong correlation in the top-ranked actions.
    }
    \label{fig:accuracy}
\end{figure}

In our study, the first-choice gestures recommended by the LLM matched 100\% with the preferred response gestures selected by humans. These findings indicate that the LLM recommendation matches the expectation of potential users (see Figure \ref{fig:accuracy}). The model's average response time during testing was 2.129 seconds.


Under the conditions of underexposed, normal, overexposed, high motion blur, moderate motion blur, low motion blur, partial subject, partial subject with off-focus, and interference from others, the LLM recommendation remained consistent. Notably, even in the completely underexposed condition, where for example the handshake gesture was nearly imperceptible to the human eye, the LLM stably recommended the "shake hand" action. However, under the high motion blur condition, the LLM was more prone to errors, with recommendation  not matching the preferences of human. These results demonstrate the LLM's potential to handle gesture recommendations across most challenging scenarios except conditions with high motion blur.

\section{Study 2: Evaluating the ABC system with Users with Limited Upper Limb Mobility}
\subsection{Participants and Procedure}


We used the same LLM model, gpt-4o-2024-05-13 Vision, as in Study 1. When saving and uploading images, we took privacy considerations into account by informing the participant in advance, and no images were retained locally. Additionally, we reviewed the LLM model company's documentation, which confirms that the image is deleted from its servers after processing, ensuring that it is neither retained nor used and thereby safeguarding the participants' privacy.

In collaboration with Organization Ory Lab Inc., which specializes in assisting individuals with disabilities to engage in social activities utilizing robotic avatars, we recruited participants who had been diagnosed with ALS or had upper limb disability, degenerative conditions that severely reduce one's mobility over time and are known to cause significant limitations to the upper limbs. In total, three participants with FSHD, one participant with Becker muscular dystrophy, and one participant with ALS expressed interest in the ABC system and agreed to participate in the study. Moreover, our collaborator with ALS who assisted in designing the initial prototype of the ABC system also volunteered to take part in the study. In total six users tested the ABC system. In Table \ref{tab:participants} we report their demographic characteristics, medical condition and mobility limitations, as well as the movements they decided to record and experiment with during the study. All participants signed informed consent prior to participation and the protocol for the study was approved by the ethics committee of [removed for anonimization]. 

\begin{table*}[htbp]
\centering
\begin{tabular}{|p{1cm}|p{2.6cm}|p{5.5cm}|p{4cm}|}
\hline
User ID  & Condition and years & Movement limitations & Recorded Movements \\ \hline
Aiko&FSHD \newline(more than 15 years)& Aiko uses a wheelchair for mobility and cannot smile or raise her arms above shoulder level, she has limited finger dexterity. She avoids gestures out of concern that they may seem unnatural. & 
1. Wave hands
2. Shake hands
3. Cheers
4. Uppercut\\ \hline

Bunji&FSHD \newline(more than 20 years)&Bunji is unable to raise his arms above his shoulders.&
1. Wave hands
2. Shake hands
3. Raise Arm
4. High five
5. Hip pop greeting
6. Clap hands\\ \hline

Chiaki&BMD \newline(more than 20 years)& Chaki uses a wheelchair for mobility, has slight upper limb impairments, and is interested in the system because of potential future difficulties with upper limb expression& 
1. Wave hands
2. Shake hands
3. Move away
4. Clap hands
5. Pray with hands clasped together\\ \hline

Daisuke&FSHD \newline(more than 15 years)&Daisuke uses a wheelchair for mobility, and cannot smile or raise his arms above his shoulders. He has limited finger dexterity and lacks the strength to assist others with handling objects.& 
1. Wave hands
2. Shake hands
3. Pass business card and  files
4. Cheers
5. Get items from high place\\ \hline

Eiji&ALS\newline(more than 11 years)&Late Stage of ALS with extremely limited mobility& 
1. Wave hands
2. Shake hands
3. Scratch an itch
4. Cheers
5. Scratch an itch
6. Turn on/off switch \\ \hline

Fumito&ALS\newline(more than 10 years)& Late Stage of ALS with extremely limited mobility&
1. Wave hands
2. Shake hands
3. Head lock
4. Scratch an itch
\\ \hline

\end{tabular}
\caption{Participants characteristics}
\Description{This table provides detailed information about six participants with various forms of muscular dystrophy or ALS, including the duration of their condition, upper limb movement limitations, and the recorded movements during the test.

Aiko (FSHD for over 15 years) and Daisuke (FSHD for over 15 years): Both use a wheelchair, cannot raise their arms or smile, and have limited finger dexterity. They avoid gestures that may seem unnatural. The recorded movements include waving hands, shaking hands, and cheering.

Bunji (FSHD for over 20 years): Unable to raise his arms, the recorded movements include waving hands, shaking hands, and clapping.

Chiaki (BMD for over 20 years): Has slight upper limb limitations and is concerned about future difficulties with upper limb expression. The recorded movements include waving hands and praying.

Eiji (ALS for over 11 years) and Fumito (ALS for over 10 years): Both in the late stage of ALS with extremely limited mobility. The recorded movements include scratching an itch and turning switches on and off.}
\label{tab:participants}
\end{table*}

Five participants used electric wheelchairs in everyday life and had varying levels of upper limb mobility. Some participants such as Aiko and Daisuke were unable to articulate facial expressions such as smiles, raise their arms, or make hand gestures like a peace sign due to limited finger dexterity. Others such as Bunji and Chaki had milder impairments and, while currently experiencing fewer difficulties, were interested in the system due to potential future challenges with body expression as a result of their progressive conditions. The final two participants Eiji and Fumito had advanced stages of ALS, resulting in severe physical impairments and a complete loss of speech, both utilized a gaze-based AAC to communicate verbally.

Despite these differences in mobility, Aiko, Bunji, Chaki, and Daisuke were able to operate the ABC system using their smartphones without needing additional assistive devices. Participants Eiji and Fumito on the other hand required assistive technologies such as PPS Switch combined with an eye-tracking device and a computer, or Switch with iPhone’s Switch Control feature, to interact with the ABC system. The study took place in our laboratory for Aiko, Bunji, Chaki, and Daisuke, whereas Eiji and Fumito chose to have the team bring the ABC system to their respective homes.

Before the experiment began, all disabled participants and the assistants of Eiji and Fumito completed questionnaires assessing their satisfaction with current communication abilities, the challenges caused by the lack of body language, and their expectations regarding the gestures the robotic arm should perform.

During the testing phase, the 6 participants, 2 assistants, and two volunteer members from our laboratory who agreed to support Aiko, Bunji, Chaki, and Daisuke were trained on how to operate the robotic arm and help record desired movements according to participants' directions. The recorded movements included standard gestures such as handshakes and waves, as well as more complex actions that participants found difficult in daily life, such as raising their arms, making a toast, handing over objects, or retrieving items from high places. Additionally, personalized gestures, such as uppercuts or headlocks, were recorded based on users’ specific interaction preferences. The recording process continued until participants were satisfied with the captured movements.

After the recording was completed, participants used the manual action retrieval function and AI recommendation feature to execute gestures, allowing them to interact with others through the ABC system. Upon completion of the tasks, participants and volunteers filled out questionnaires, including the System Usability Scale (SUS) \cite{grier2013system}, NASA Task Load Index (NASA TLX) \cite{hart2006nasa}, Sense of Agency \cite{tapal2017sense} and rated again their satisfaction with communication abilities using accounting for the use of the robot arm. Finally, participants underwent brief semi-structured interviews to provide feedback on their experience with the system. User testing sessions were video recorded and interviews were later transcribed for analysis. Videos and transcripts were collectively analyzed by the research team using affinity diagrams \cite{lucero2015using}. Data from questionnaires was charted and summarized.
\begin{figure}
    \centering
    \includegraphics[width=1\linewidth]{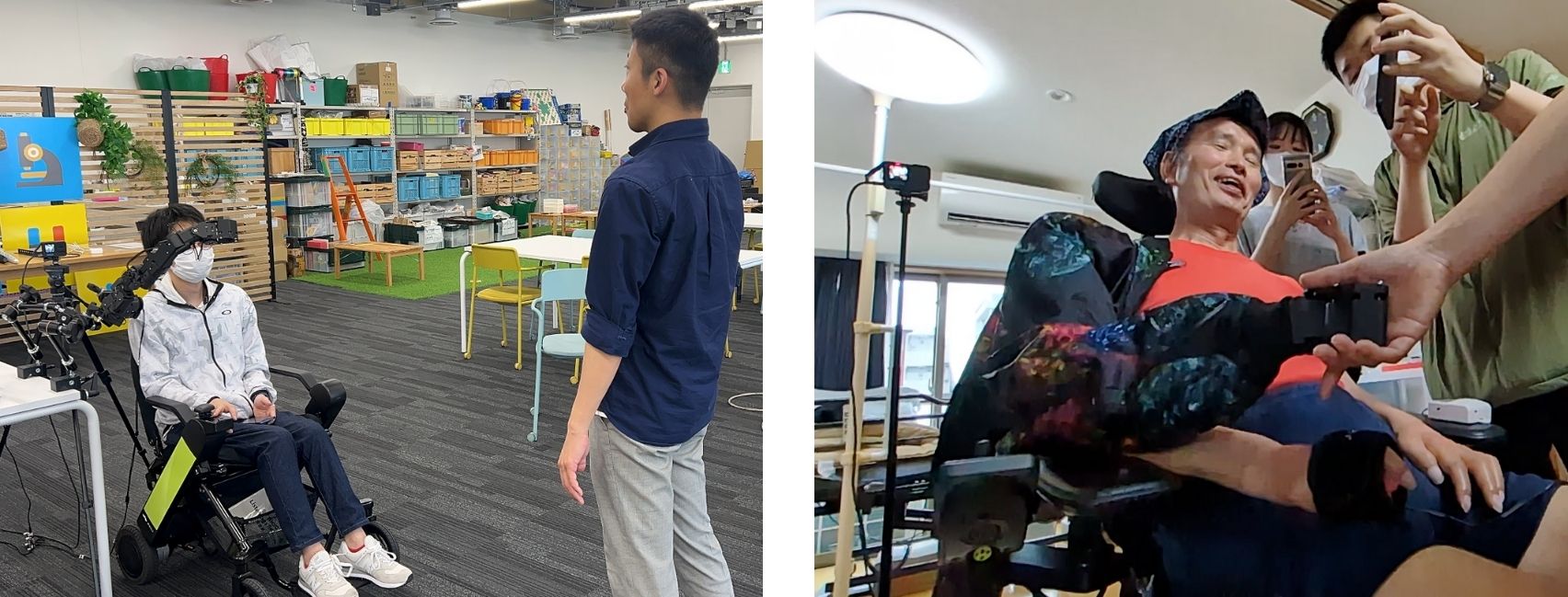}
    \caption{ABC In-depth User test}
    \Description{This image shows two scenes related to the testing of ABC system for individuals with physical disabilities.
    
    The left side depicts a participant in a wheelchair operating a robotic arm during a test, with an interactor standing in front of them. The right side shows another participant, also in a wheelchair, using the robotic arm with the assistance of others.}
    \label{fig:enter-label}
\end{figure}
\subsection{Findings}
\subsubsection{Quantitative Findings}

In the SUS scale, the six participants reported an average SUS score of $ 90 \pm 2.6 $, which indicates high usability. On the NASA TLX scale, the average RAW TLX score for the six participants was $ 7.33 \pm 5.3 $ indicating a low workload. In the Sense of Agency scale, the average Sense of Positive Agency (SoPA) score, which indicates the feeling of control over the ABC on a 7-point scale, was $ 4 \pm 0.7 $, while the average Sense of Negative Agency (SoNA) score, indicating perceived lack of control of the ABC on a 7-point scale, was $ 1.01 \pm 0.8 $. These results indicate that the system performed well in terms of user experience, with most users expressing high satisfaction with the system's usability and functionality. Compared to others participants, Eiji reported lower satisfaction as he felt that while the AI system was effective, using manual searches could be cumbersome when using a gaze-based interface. A paired-sample T-test was conducted to compare users' self-evaluations before and after using the ABC system. The mean self-evaluation score before the experiment was ((M = 2.83, SD = 0.98), and after the experiment, the mean score increased to (M = 4.33, SD = 0.82). There was a statistically significant difference in the scores before and after the experiment, $t(5) = 2.67$, $p = 0.045$. These results, albeit from a very small group of participants which suggest careful interpretation, seem to indicate that the ABC system positively impacted users' self-perceived ability to express themselves through body language.

\subsubsection{Qualitative Findings}\mbox{}

\textbf{ABC and AAC}
\newline
Participants found the ABC system to be an effective communication tool that was similar to, and could complement, an AAC device. For example, Aiko explained that although she retains some upper limb movement, she rarely uses her hands for communication or expressive gestures because people often misinterpret her gestures (e.g. a peace sign as commonly posed in Japan) or she feels self‐conscious about her movements.

\emph{"Due to difficulties raising my arm in daily life, I often feel embarrassed and end up avoiding movements like waving. This can make me feel a bit down. However, being able to wave with the robotic arm could brighten my mood and make moments like saying goodbye to friends feel more cheerful."} Aiko

Her challenges echoed those in the literature regarding non-standard speech that is often misunderstood \cite{dai2022designing}. Aiko noted that with her specifications met and the volunteer’s support, the ABC system boosted her confidence in social gestures.

Both Eiji and Fumito saw the ABC not merely as an AAC substitute but as a complementary tool. In one-on-one conversations with a patient partner, delays in AAC input were manageable, yet in group settings the pace left them feeling excluded. (\textit{"While I’m entering text using eye-tracking input, the conversation often moves on to the next topic, which makes me feel frustrated."} Fumito). They explained that a low AAC volume can hinder being heard; however, by clicking a pre-selected movement, they could raise the robotic arm to attract attention and participate more effectively. Moreover, participants felt that the arm’s human-like form helped others interpret their intentions.

Although the ABC system alleviated some AAC-related issues, it shared limitations such as delays in social exchanges. Both Eiji and Bunji remarked that manually searching for a desired movement was cumbersome, especially when many movements were available. Even the AI-recommended process required about 5 seconds to process an image and suggest a movement, which could cause awkward delays that Aiko and Daisuke found concerning.

\textbf{My arm, my body language}
\newline
All participants expressed a strong sense of embodiment with the ABC system’s arm. Assistants for Eiji and Fumito confirmed that the arm’s movements appeared natural, as if produced by the user’s own body. They agreed that for the ABC to feel like a part of one’s body, its movements must match the user’s expected body language. Participants and assistants invested time in experimenting with, recording, and refining movements until they were satisfied. Explaining movement details and desired joint speeds was challenging to do verbally, leading Daisuke to wish for a better method to convey this information clearly.

\emph{"The current setup, where you communicate your desired movements to a caregiver and they input them, it’s sometimes hard to convey your intentions with just words. AI analysis could help supplement this, so it would be good to have such a feature to explain and show what I mean."} Daisuke

Bunji and Aiko mentioned that, while they normally do not use personal assistants, they could not activate the robotic arm to create new gestures by themselves. They would prefer a digital interface that allows them to design and test animations on screen before sending them to the arm. Chaki, who still had good residual upper limb mobility, could move the arm but struggled with recording motions accurately because of dexterity issues; he wondered if he could record movements via video instead of manual input.

Finally, Eiji, Bunji, and Daisuke noted that in daily life, arms serve both communicative and functional roles. They argued that, given the arm’s flexible design, the ABC system should also support everyday tasks.

\emph{"Our arms have two functions: one is body language for social interaction, and the other is performing precise tasks. So far, there hasn’t been a functional assistive device for people with arm disabilities that can be used for language and tasks. I see great potential in such a device."} (Bunji)

\textbf{Need for reciprocity}
\newline
Participants and their assistants felt that the ABC system’s key benefit was its ability to help those with upper limb limitations engage more actively in social exchanges. They noted that many social gestures—such as waves, handshakes, hugs, or even business card exchanges—are reciprocal by nature. People with restricted mobility might receive these gestures but often cannot reciprocate. Fumito’s assistant summarized this need:

\emph{"Fumito is passionate about running and enjoys challenges. After completing relay marathons with friends or finishing a 10km run together, or succeeding in other challenges, Fumiko typically expresses joy through a big smile. While others might use high-fives, hugs, or handshakes to share their happiness, Fumito’s way of expressing joy is through their broad smile. Joyful moments often involve fleeting surges of emotion, and Fumito believes that being able to share these moments with gestures like high-fives or hugs would help deepen the sense of empathy and connection with others."} (Fumito's Assistant)

Both Chaki and Fumito suggested that integrating a tactile feedback interface could transform handshakes or hugs by providing proprioceptive feedback. This feedback would make users feel as if they were truly performing the gesture rather than merely triggering a response.

However, Eiji pointed out a flaw in the AI selection mode, which depends on the interaction partner initiating contact—such as offering a handshake—because the system uses the partner’s image to suggest a response. This means the participant cannot start an exchange with the AI mode and must rely on the manual search option, which can be cumbersome with adaptive interfaces. Eiji and Fumito proposed that quick triggers—like individual switches or specific eye commands—be linked to common gestures for more spontaneous interactions. Overall, participants were excited by the ABC system’s potential to change how others perceive them during social interactions.

\emph{"In daily interactions, people tend to treat me differently due to my physical condition. However, after using the robotic arm, I felt that others would ignore my physical limitations and interacted with me normally."} - Aiko.

\section{Discussion}

Limitations in physical as well as verbal expression can have a severe negative impact on social participation\cite{lyons2017labels}. Although AAC devices can help support one's ability to engage in verbal communication, there are still limited solutions which can assist individuals with their nonverbal communication\cite{valencia2021aided, valencia2021co, sobel2017exploring}. Our work centred on the development and testing of an ABC system to support body expressions during social exchanges for people with severe mobility limitations. Insights gathered from participants offered important guidance for the improvement of the system but also stimulated reflections around the potential of expansion and integration with existing AAC technologies to create more  Augmented Comprehensive Communication technologies (ACCs) to meaningfully support communication needs in a more holistic manner. Feedback from the user studies also helped to consider opportunities and drawbacks associated with the use of AI as part of these systems to augment communication. In the following subsections, we attempt to unpack both these aspects. 


\subsection{From targeted AACs and ABCs to holistic ACCs}
When we are feeling sad and upset, would a kind word mean more to us than a hug? When facing a tough challenge is a hand on the shoulder more effective than a heartfelt encouragement? These questions cannot be answered without context because the preferred option, if exists, depends on the individuals, their relationships and the circumstances at that particular moment. The reality is that both verbal and nonverbal communication represents a meaningful and essential component of most if not all, our social exchanges. Yet in the development of technology to aid communication, we tend to treat these two aspects as separate. AAC devices can allow participants to continue to engage in verbal communication when they have lost the ability to speak \cite{valencia2020conversational, kane2017times}. Practices such as voice banking can also enable people, especially those with degenerative conditions such as ALS, to collect samples of their own voice, to be able to train a personalized voice model to import on their AAC at a later time \cite{bunnell2017modeltalker}. 

Although AAC-enabled verbal communication can support individuals in conveying reach and detailed information through the content of the message and the intonation of the voice, it misses the subtle qualities and embodied expressions of nonverbal communication. In previous research, we noticed how embodied agents such as the Sidekicks by Valencia et al \cite{valencia2021co} or badges like the ones created by Sobel et al \cite{sobel2017exploring} are intended as having a supporting role for the AAC seen as the main tool for communication. However, during our study, we noticed how many of the participants did not necessarily the gestures they created in a situation where verbal conversation was also present. Verbal and non-verbal communication often coexist, but they can also be used separately and our ABC was designed to enable bodily expressions with or without the use of associated speech. 

However, if verbal and nonverbal communication are so deeply intertwined, why do we design communication systems that support either one or the latter, but not the two of them together? Although it is in theory possible for one person to use both an AAC and an ABC, in fact, both Eiji and Fumito in our second study did so, it can be extremely challenging to control them at the same time. This is especially important because individuals likely to benefit from using both systems have severe disabilities that reduce the possibilities for input modalities. Moreover, evidence from our study and previous research points out how those who use eye-gaze for AAC or ABC control, already experience challenges in social exchanges due to the inherent difficulties and delays associated with gaze-base devices \cite{valencia2021aided,kane2017times}. Adding complexity for the sake of more functionality would only exacerbate the problem. 

On the other hand, a combined ACC system could offer a more flexible solution to potential users, allowing individuals to decide whether they want to prioritize verbal or not verbal communication depending on the situation, but also explore opportunities to combine them by, for example adjusting the posture of the ABC or generating movements that match the content and tone of verbal messages that participants are speaking through the AAC. These strategies have been explored in the field of robotics as a way to increase the appeal and perceived complexity of personality \cite{zabala2021expressing, lai2024intuitive} but, to our knowledge they have never been investigated as a way to support comprehensive communication needs for people with both impaired speech and limited mobility.

\subsection{Challenges and Opportunities of AI for ABC systems}

Our ABC system featured the integration of an LLM AI-based image recognition that supports the user's ability to select a movement from a pre-recorded set based on the degree of match with the action performed by a conversational partner which is captured by a camera. The user study uncovered some of the limitations connected to the way the system operates, primarily the inability of the user to initiate an exchange using the same system. Moreover, it struggled with recommending unique greeting styles which some of the participants in our second study created, such as head locking or uppercut movement. Additionally, likely due to the model's emphasis on promoting positive content, it is unlikely to recommend actions that convey negative, impolite, or passive emotions, which limits its ability to meet the unique needs of users. This is similar to some of the issues witnessed with AACs censoring people and preventing them from swearing, regardless of their personal wishes, due to settings embedded in a particular application \cite{kane2017times}. 

Alternatives to issues with initiations could be addressed by incorporating mechanisms that could trigger recommendations based on more general actions, like a person approaching activating a selection amongst a subset of movements all labelled as greetings. However, care should be considered when adopting mechanisms that require the LLM AI-based Image recognition API due to potential concerns about privacy and continuous capturing of video data from passersby without their explicit consent. This is also something that we discussed as a team in relation to future deployment of the current system. 
One solution to this issue, which could also address some of the challenges around personalization, would be to fine-tune and create personalized models for each individual which could also run locally on a device. This approach is similar to the one used by recent automatic speech recognition technologies for users with non-standard speech \cite{ayoka2024enhancing}. Such an approach would address both concerns about privacy and personalization but would require a significant investment of time and effort from the user, especially in the early stages, to appropriately train the image recognition-based recommendation model and complement more generic LLMs.

Finally, we point to two potential additional uses for AI technologies based on the findings that emerged from our study. First, as pointed out by Aiko and Bunji, it is important for users to necessarily have to rely on assistants to create new movements for the ABC. To address this challenge it is possible to incorporate inverse kinematics (IK) functionality into the ABC to enable the creation of animations from a dedicated application. However, such a task could be challenging for someone with reduced upper limb mobility as kinematic movement design often requires small adjustments to produce satisfactory results. The incorporation of a generative AI interface could enable users to create the desired movements by repeated prompts or analogy. Secondly, as noted by Chaki, people who still have residual upper limb movements, but are concerned about losing mobility due to degenerative conditions, could be supported in creating personalized "movement banks" for ABC systems to use in the future, similarly to the practice of voice banking for AAC users \cite{bunnell2017modeltalker}.

\section{Limitations and Future Works}



During user testing, almost all participants expressed a desire to use the robotic arm for object interactions or functional tasks, such as retrieving items from high places, controlling switches, and passing or receiving objects. 

In our early experiments, we previously attempted to let the LLM directly generate motor values for the robotic arm, but errors occurred, leading to uncontrolled movements. For safety reasons, we now only allow users' assistants to record the robotic arm's actions, which are then replayed by the user. This ensures control but  results in limited action variety. While users can record nearly infinite actions, the robotic arm cannot autonomously execute animations nor perform complex functional tasks based on environmental cues. In the future, we aim to incorporate motion-matching algorithms from the gaming industry \cite{holden2020learned}, along with path-planning algorithms, to guide the robotic arm in executing more complex movements.

Although current testing has focused on everyday interactions, the studies took place in users' homes and relatively stable lab environments. We have not yet tested long-term use in diverse real-life settings. In the future, we plan to explore user feedback and evaluation of the ABC system based on long-term daily use in different environments.
\section{Conclusion}



In this work, we propose an ABC system that can enhance the physical and social interaction capabilities of individuals with upper limb mobility impairments. By integrating a robotic arm with kinetic memory, LLM AI-powered image recognition, and adaptable control mechanisms, the ABC system enables users to perform meaningful gestures and actions that were previously difficult or impossible to achieve. We first compared the preferences of the LLM with those of 14 participants in selecting response actions and measured the LLM's accuracy in action selection under different manually altered image conditions. We then conducted a mixed-method in-depth user test with six users with upper limb impairments. Our results show that the system can expand users' expressiveness in social contexts.

We found the LLM demonstrated a high similarity to human tendencies in responding to body movements based on images, as well as high accuracy under different conditions. Additionally, disabled users acknowledged the system's ability to foster more personalized and interactive communication, allowing users to actively initiate and engage in everyday social interactions. 

Our research contributes to the growing body of work on using robotic arms to support both functional and social interaction for individuals with severe mobility impairments. By continuing to refine the design and functionality of systems like ABC, we can further empower users to better express themselves and engage with the environment around them.

\begin{acks}
This work was supported by JST Moonshot R\&D Program "Cybernetic being" Project (Grant number JPMJMS2013). 
\end{acks}

\bibliographystyle{plain}  
\bibliography{references}

\end{document}